\documentclass[aps,prd,onecolumn,groupedaddress,showpacs,nofootinbib,amssymb]{revtex4}
\usepackage[dvips]{graphicx}
\usepackage{amssymb}
\usepackage{amsmath}
\usepackage{graphicx}
\usepackage{amsfonts}
\usepackage{bm}

\begin{document}

\title{Viable Mimetic $F(R)$ Gravity Compatible with Planck Observations}

\author{S.D. Odintsov$^{1,2,6}$, V.K. Oikonomou$^{4,5}$}
\affiliation{$^{1)}$Instituci\`{o} Catalana de Recerca i Estudis Avan\c{c}ats
(ICREA),
Barcelona, Spain \\
$^{2)}$Institut de Ciencies de l'Espai (CSIC-IEEC), Campus UAB,
Campus UAB, Carrer de Can Magrans, s/n 08193 Cerdanyola del Valles, Barcelona, Spain\\
$^{4)}$  Tomsk State Pedagogical University,  634061 Tomsk, Russia \\
$^{5)}$  Lab. Theor. Cosmology, Tomsk State University of Control Systems\\
and Radioelectronics, 634050 Tomsk, Russia (TUSUR)\\
$^{6)}$  National Research Tomsk State University,  634050 Tomsk }

\begin{abstract}
Using the mimetic matter approach, we study $F(R)$ gravity
 with scalar potential and Lagrange multiplier constraint. As we demonstrate, for a given $F(R)$ gravity and for suitably chosen mimetic potential, it is possible to realize inflationary cosmology consistent with Planck observations. We also investigate the de Sitter solutions of the mimetic
 $F(R)$ theory and  study the stability of the solutions, when these exist, towards linear perturbations, with the unstable solutions,
 which can provide a mechanism for graceful exit from inflation. Finally, we describe a reconstruction method which can yield the $F(R)$ gravity that can generate realistic inflationary cosmological evolution, given the mimetic potential and the Hubble rate.
\end{abstract}


\maketitle

\section{Introduction}

Our Universe's current evolution status is determined by the contribution of ordinary matter with $\Omega_m\sim4.9\%$, cold dark matter with $\Omega_{DM}\sim
26.8\%$ and also dark energy with $\Omega_{DE}\sim 68.3\%$, with the dark energy contribution controlling the late-time acceleration, verified observationally in the late 90's \cite{latetimeobse}. A consistent description of late-time acceleration is provided by the modified theories of gravity (see reviews,\cite{reviews1}). There is number of modified theories of gravity that can model quite successfully dark energy (for discussion on dark energy, see\cite{reviews2}). On the contrary, for cold dark matter very few are known, apart from the fact that this form of matter modifies the galactic curves. In the literature, it is common to consider dark matter as a particle that has no-interaction with luminous matter. In the context of supersymmetric modifications of the Standard Model, one particle that is a potential candidate for describing cold dark matter, is the neutralino \cite{shafi}, but there exist also other particles that could describe dark matter, see for example \cite{darkmatter} and references therein. Apart from the particle nature of cold dark matter, there exists a recent alternative description to cold dark matter, provided by the so-called mimetic modification of General Relativity, proposed by Chamseddine and Mukhanov \cite{mukhanov1} and further studied in \cite{mukhanov2,Golovnev}. In their paper \cite{mukhanov1}, Chamseddine and Mukhanov isolated the conformal degree of freedom of Einstein-Hilbert gravity in a covariant way, and in the resulting theory, the physical metric is defined with the account of an auxiliary scalar field, which appears through its first derivatives. In effect, the conformal degree of freedom becomes dynamical, even in the absence of any matter fluids, and more importantly this conformal degree of freedom can mimic cold dark matter. A natural extension of the mimetic Einstein-Hilbert gravity is provided by mimetic $F(R)$  gravity \cite{NO2}, since these theories could potentially have the attributes of the mimetic dark matter approach, but it is possible to also describe early-time and late-time acceleration within the same theoretical framework \cite{NO3}.

In this letter we shall study Jordan frame mimetic $F(R)$  gravity, with the mimetic dark matter part having also a potential term. For some informative reviews on $F(R)$ theories we refer to \cite{reviews1}, and for some early but worthy to mention studies, see \cite{barrowearly}. We shall use the Lagrange multiplier approach \cite{CMO} to modified gravity, so that we achieve a consistent incorporation of the mimetic matter potential, in the Jordan frame $F(R)$ framework. Our aim is to provide a reconstruction technique for obtaining the mimetic potential which generates a cosmological evolution compatible with the recent Planck data \cite{planck}, by using viable $F(R)$ theory, with viable meaning that they satisfy the local and large scale constraints posed on these by observations \cite{reviews1}. We shall mainly work in the Jordan frame and in order to have a better command on the cosmological evolution we shall use the $e$-folding number $N$ as a variable, and we shall express all the involved to the calculation quantities in terms of this variable. Also,
 we shall  briefly provide a second reconstruction technique, which enables us to find which $F(R)$ gravity can generate a specific cosmological evolution with a specific mimetic potential. Finally, we investigate the existence of unstable de Sitter solutions for some $F(R)$ gravities with mimetic potential.

\section{Consistent  Cosmologies from Lagrange-multiplier Mimetic $F(R)$ Gravity}

In the context of mimetic approach, the conformal symmetry is considered as an internal degree of freedom and it is not violated. The physical metric $g_{\mu \nu}$, is expressed in terms of an auxiliary metric $\hat{g}_{\mu \nu}$ and of an auxiliary scalar field $\phi$, in the following way:
\begin{equation}\label{metrpar}
g_{\mu \nu}=-\hat{g}^{\mu \nu}\partial_{\rho}\phi \partial_{\sigma}\phi \hat{g}_{\mu \nu}\, .
\end{equation}
Instead of considering the variation of the gravitational action with respect to the physical metric $g_{\mu \nu}$, one can consider variation with respect to the auxiliary metric $\hat{g}_{\mu \nu}$ and with respect to $\phi$. From Eq. (\ref{metrpar}) it easily follows that,
\begin{equation}\label{impl1}
g^{\mu \nu}(\hat{g}_{\mu \nu},\phi)\partial_{\mu}\phi\partial_{\nu}\phi=-1\, .
\end{equation}
The parametrization (\ref{metrpar}) is invariant under the Weyl transformation $\hat{g}_{\mu \nu}=e^{\sigma (x)}g_{\mu \nu}$, and the auxiliary metric $\hat{g}_{\mu \nu}$ does not appear explicitly in the action. We shall assume that the physical metric $g_{\mu \nu}$ is a flat Friedmann-Robertson-Walker (FRW) metric with line element,
\begin{equation}\label{frw}
ds^2 = - dt^2 + a(t)^2 \sum_{i=1,2,3}
\left(dx^i\right)^2\, ,
\end{equation}
where the parameter $a(t)$ denotes the scale factor, and the  scalar  $\phi$ depends only on time. For the flat FRW background, the Ricci scalar has the form $R=6\left (\dot{H}+2H^2 \right )$, where $H(t)$ is the Hubble rate. The Jordan frame gravitational action for the mimetic $F(R)$ theory with scalar potential $V(\phi )$ and Lagrange multiplier $\lambda (\phi )$, is the following,
\begin{equation}\label{actionmimeticfraction}
S=\int \mathrm{d}x^4\sqrt{-g}\left ( F\left(R(g_{\mu \nu})\right )-V(\phi)+\lambda \left(g^{\mu \nu}\partial_{\mu}\phi\partial_{\nu}\phi +1\right)\right )\, ,
\end{equation}
where for simplicity, no extra matter  is present. By varying the action (\ref{actionmimeticfraction}), with respect to the physical metric $g_{\mu \nu}$, we can easily obtain,
\begin{align}\label{aeden}
& \frac{1}{2}g_{\mu \nu}F(R)-R_{\mu \nu}F'(R)+\nabla_{\mu}\nabla_{\nu}F'(R)-g_{\mu \nu}\square F'(R)\\ \notag &
\frac{1}{2}g_{\mu \nu}\left (-V(\phi)+\lambda \left( g^{\rho \sigma}\partial_{\rho}\phi\partial_{\sigma}\phi+1\right) \right )-\lambda \partial_{\mu}\phi \partial_{\nu}\phi =0 \, .
\end{align}
Varying the action of Eq. (\ref{actionmimeticfraction}), with respect to the auxiliary scalar $\phi$, we obtain,
\begin{equation}\label{scalvar}
-2\nabla^{\mu} (\lambda \partial_{\mu}\phi)-V'(\phi)=0\, ,
\end{equation}
where the prime denotes differentiation with respect to the auxiliary scalar $\phi$. Upon varying the action (\ref{actionmimeticfraction}) with respect to $\lambda$, we get,
\begin{equation}\label{lambdavar}
g^{\rho \sigma}\partial_{\rho}\phi\partial_{\sigma}\phi=-1\, ,
\end{equation}
which is identical to Eq. (\ref{impl1}). Assuming the flat FRW background of Eq. (\ref{frw}), and also that the scalar field depends only on the cosmic time $t$, the equations of motion (\ref{aeden}), (\ref{scalvar}) and (\ref{lambdavar}) are written as follows,
\begin{equation}\label{enm1}
-F(R)+6(\dot{H}+H^2)F'(R)-6H\frac{\mathrm{d}F'(R)}{\mathrm{d}t}-\lambda (\dot{\phi}^2+1)+V(\phi)=0\, ,
\end{equation}
\begin{equation}\label{enm2}
F(R)-2(\dot{H}+3H^2)+2\frac{\mathrm{d}^2F'(R)}{\mathrm{d}t^2}+4H\frac{\mathrm{d}F'(R)}{\mathrm{d}t}-\lambda (\dot{\phi}^2-1)-V(\phi)=0\, ,
\end{equation}
\begin{equation}\label{enm3}
2\frac{\mathrm{d}(\lambda \dot{\phi})}{\mathrm{d}t}+6H\lambda \dot{\phi}-V'(\phi)=0\, ,
\end{equation}
\begin{equation}\label{enm4}
\dot{\phi}^2-1=0\, ,
\end{equation}
where the ``dot'' in all the above equations denotes differentiation with respect to the cosmic time $t$, while the ``prime'' in Eqs. (\ref{enm1}) and (\ref{enm2}) denotes differentiation with respect to the scalar curvature, while in Eq. (\ref{enm3}) denotes differentiation with respect to the scalar field $\phi$. From Eq. (\ref{enm3}), it follows that the scalar field $\phi$ can be identified with the cosmic time $t$ (the same identification was possible in the Einstein-Hilbert mimetic gravity, as an implication of Eq. (\ref{impl1}), see \cite{mukhanov1}). Since $t=\phi$, Eq. (\ref{enm2}) can be written in the following way,
\begin{equation}\label{sone}
F(R)-2(\dot{H}+3H^2)+2\frac{\mathrm{d}^2F'(R)}{\mathrm{d}t^2}+4H\frac{\mathrm{d}F'(R)}{\mathrm{d}t}-V(t)=0\, ,
\end{equation}
from which it easily follows that the scalar field potential $V(\phi=t)$ that generates the Hubble rate $H(t)$, for a specific $F(R)$ gravity, is equal to,
\begin{equation}\label{scalarpot}
V(\phi=t)=2\frac{\mathrm{d}^2F'(R)}{\mathrm{d}t^2}+4H\frac{\mathrm{d}F'(R)}{\mathrm{d}t}+F(R)-2(\dot{H}+3H^2)\, .
\end{equation}
Upon combining Eqs. (\ref{scalarpot}) and (\ref{enm1}), we obtain the analytic form of the Lagrange multiplier, as a function of the cosmic time $t$, which reads,
\begin{equation}\label{lagrange}
\lambda (t)=-3 H \frac{\mathrm{d}F'(R)}{\mathrm{d}t}+3 (\dot{H}+H^2)-\frac{1}{2}F(R)\, .
\end{equation}
Then, we have at hand a reconstruction method with which we can find which scalar potential and Lagrange multiplier can generate a specific Hubble rate, for a given arbitrarily chosen $F(R)$ gravity.

The novelty of this reconstruction method we propose is that, in principle we can choose a viable $F(R)$ gravity and a Hubble rate that is compatible with the Planck data and find the corresponding mimetic scalar potential that generates the evolution, for the given $F(R)$ gravity. In order to exemplify our method, we shall use an illustrative example by choosing an arbitrary $F(R)$ gravity. Consider for example the Jordan frame cubic $F(R)$ gravity,
\begin{equation}\label{staro}
F(R)=R+d R^3\, ,
\end{equation}
with arbitrary parameter $d$. Our aim is to find which scalar field potential $V(\phi=t)$ reproduces a cosmology compatible with the observations, for the $F(R)$ gravity chosen as in Eq. (\ref{staro}). We shall use the slow-roll parameters and observational indices found in Ref. \cite{sergeislowroll} (see also \cite{muk}).
As was shown in  \cite{sergeislowroll}, the $F(R)$ gravity can be treated as a perfect fluid, so the slow-roll parameters and the observational indices can be calculated in terms of the Hubble rate easily. In our case, the procedure is the same, so we do not reproduce it here. Hence, we shall calculate the slow-roll parameters, for a viable cosmological expansion and then determine the potential $V(t=\phi)$, which generates this expansion. In order to have a better command on the resulting equations, we shall use the $e$-folding number $N$, instead of the cosmic time $t$. Using the following transformation rules for the derivatives with respect to the cosmic time and to the $e$-folding number,
\begin{equation}\label{transfefold}
\frac{\mathrm{d}}{\mathrm{d}t}=H(N)\frac{\mathrm{d}}{\mathrm{d}N},{\,}{\,}{\,}
\frac{\mathrm{d}^2}{\mathrm{d}t^2}=H^2(N)\frac{\mathrm{d}^2}{\mathrm{d}N^2}+H(N)\frac{\mathrm{d}H}{\mathrm{d}N}\frac{\mathrm{d}}{\mathrm{d}N}\, ,
\end{equation}
and following \cite{sergeislowroll}, the slow-roll indices can be written in terms of the $e$-folding number $N$, as follows,
\begin{align}\label{hubbleslowrollnfolding}
&\epsilon=-\frac{H(N)}{4 H'(N)}\left(\frac{\frac{H''(N) }{H(N)}+6\frac{H'(N) }{H(N)}+\left(\frac{H'(N)}{H(N)}\right)^2}{3+\frac{H'(N)}{H(N)}}\right)^2
\, ,\\ \notag &
\eta=-\frac{\left(9\frac{H'(N)}{H(N)}+3\frac{H''(N)}{H(N)}+\frac{1}{2}\left( \frac{H'(N)}{H(N)}\right)^2-\frac{1}{2}\left( \frac{H''(N)}{H'(N)}\right)^2+3 \frac{H''(N)}{H'(N)}+\frac{H'''(N)}{H'(N)}\right)}{2\left(3+\frac{H'(N)}{H(N)}\right)}
\end{align}
where the prime denotes differentiation with respect to the $e$-folding number $N$. The spectral index of primordial curvature perturbations and the scalar-to-tensor ratio $r$, are equal to,
\begin{equation}\label{indexspectrscratio}
n_s\simeq 1-6 \epsilon +2\eta,\, \, \, r=16\epsilon \, .
\end{equation}
Notice that these relations are valid, so long as during inflation (which means for most values of $N$), the slow-roll indexes are $\epsilon,\eta \ll 1$, so this formalism is valid only if the slow-roll approximation is assumed to hold true.

Having these at hand, we can choose a cosmological evolution compatible with the recent Planck data and find the corresponding potential $V(N)$ that generates this evolution, for the $F(R)$ function being of the form (\ref{staro}). As we already mentioned, we shall be interested in the spectral index of primordial curvature perturbations $n_s$ and the scalar-to-tensor ratio $r$, which are constrained by the recent Planck data \cite{planck}, as follows,
\begin{equation}\label{constraintedvalues}
n_s=0.9644\pm 0.0049\, , \quad r<0.10\, .
\end{equation}
Let us now exemplify our method by using some characteristic examples. Consider for example the class of models of cosmological evolution with Hubble rate of the form,
\begin{equation}\label{hub1}
H(N)=\left(-G_0\text{  }N^{\beta }+G_1\right)^b\, ,
\end{equation}
where $G_0$, $G_1$, $\beta$ and $b$ arbitrary real numbers in general. Since for negative values of $G_0$ and $G_1$, the Hubble rate might become negative, we choose $b$ to be of the form, $b=\frac{2n}{2m+1}$, with $m$ and $n$ positive integers chosen in such a way so that $b<1$. The slow roll parameters for the Hubble rate (\ref{hub1}) are equal to,
\begin{align}\label{hubpar1}
& \epsilon=N^{1-\beta } \Big{(}G_1-G_0 N^{\beta }\Big{)}\Big{(}3-\frac{b G_0 N^{-1+\beta } \beta }{G_1-G_0 N^{\beta }}\Big{)}^{-2}(4 b G_0 \beta )^{-1} \Big{(}-\frac{6 b G_0 N^{-1+\beta } \beta }{G_1-G_0 N^{\beta }}+\frac{b^2 G_0^2 N^{-2+2 \beta } \beta ^2}{\Big{(}G_1-G_0 N^{\beta }\Big{)}^2} \\ \notag &
+\Big{(}G_1-G_0 N^{\beta }\Big{)}^{-b} \Big{(}-b G_0 N^{-2+\beta } \Big{(}G_1-G_0 N^{\beta }\Big{)}^{-1+b} (-1+\beta ) \beta  \\ \notag &
+(-1+b) b G_0^2 N^{-2+2 \beta } \Big{(}G_1-G_0 N^{\beta }\Big{)}^{-2+b} \beta ^2\Big{)}\Big{)}^2 \, .
\end{align}
with regards to the parameter $\epsilon$, while the parameter $\eta$ reads,
\begin{align}\label{etaparameterversion1}
& \eta=-\frac{G_1^2 (-1+\beta ) (-3+6 N+\beta )+G_0^2 N^{2 \beta } \left(3-10 b \beta +8 b^2 \beta ^2+6 N (-1+4 b \beta )\right)}{4 N \left(G_1-G_0 N^{\beta }\right) \left(3 G_1 N-G_0 N^{\beta } (3 N+b \beta )\right)}\\ \notag &
+\frac{-2 G_0 G_1 N^{\beta } (3 N (-2+\beta +4 b \beta )+(-1+\beta ) (-3+(-1+5 b) \beta ))}{4 N \left(G_1-G_0 N^{\beta }\right) \left(3 G_1 N-G_0 N^{\beta } (3 N+b \beta )\right)}
\end{align}
In view of Eqs. (\ref{hubpar1}) and (\ref{etaparameterversion1}), the observational indices of Eq. (\ref{indexspectrscratio}) become,
\begin{align}\label{observa}
& n_s=\frac{1}{2 N \Big{(}G_1-G_0 N^{\beta }\Big{)} \Big{(}-3 G_1 N+G_0 N^{\beta } (3 N+b \beta )\Big{)}^2}\times \\ \notag &
\Big{(}3 G_1^3 N \Big{(}-3+6 N (1+N)+4 \beta -6 N \beta -\beta ^2\Big{)}-G_0^3 N^{3 \beta } (9 N (-1+2 N (1+N)) \\ \notag &
 +48 b N^2 \beta +2 b^2 (-1+13 N) \beta ^2+4 b^3 \beta ^3\Big{)}-G_0 G_1^2 N^{\beta } \Big{(}54 N^3+2 b (-1+\beta ) \beta ^2+3 N (-1+\beta ) (9+\beta ) \\ \notag &  +6 N^2 (9-6 \beta +8 b \beta )\Big{)}+G_0^2 G_1 N^{2 \beta } \Big{(}54 N^3+2 b (1+b) (-1+\beta ) \beta ^2 \\ \notag &
 +6 N^2 (9-3 \beta +16 b \beta )+N \Big{(}-27+2 \beta  \Big{(}6+3 \beta +13 b^2 \beta \Big{)}\Big{)}\Big{)}\Big{)} \, .
\end{align}
By choosing the free parameters $G_0$, $G_1$, $\beta$ and $b$ as follows,
\begin{equation}\label{defparam}
G_0=0.00005,\, \, \, G_1=3000, \, \, \,\beta=3, \, \, \, b=\frac{8}{9}\, ,
\end{equation}
the corresponding spectral index and scalar-to-tensor ratio for nearly $e$-folds ($N=60$), are equal to,
\begin{equation}\label{scalindex}
n_s\simeq 0.966157, \, \, \, r\simeq 0.000258947\, ,
\end{equation}
which are compatible with the observational constraints (\ref{constraintedvalues}). Hence, since the cosmology generated by (\ref{hub1}) is compatible with observational data, we can straightforwardly find the scalar potential that can generate such a cosmology. By introducing the function $G(N)$, which is equal to $G(N)=H(N)^2$, we can easily see that the Ricci scalar $R$, is related to the function $G(N)$ as follows (see also \cite{sergeirecon}),
\begin{equation}\label{cnc}
R(N)=3G'(N)+12G(N)\, .
\end{equation}
 By combining relations (\ref{hub1}), (\ref{staro}) and (\ref{cnc}), and by substituting in (\ref{scalarpot}), we obtain the scalar potential $V(N)$, which is equal to,
\begin{align}\label{explicitpot1}
& V(N)=6 \left(\mathcal{K}(N)\right)^{2 b}+12 d \left(\mathcal{K}(N)\right)^{2 b}+432 d \left(\mathcal{K}(N)\right)^{4 b}+1728 d \left(\mathcal{K}(N)\right)^{6 b} \\ \notag & -4 b G_0 N^{-1+\beta } \left(\mathcal{K}(N)\right)^{-1+2 b} \beta
-216 b d G_0 N^{-1+\beta } \left(\mathcal{K}(N)\right)^{-1+4 b} \beta-2592 b d G_0 N^{-1+\beta } \left(\mathcal{K}(N)\right)^{-1+6 b} \beta \\ \notag &  +1296 b^2 d G_0^2 N^{-2+2 \beta } \left(\mathcal{K}(N)\right)^{-2+6 b} \beta ^2-216 b^3 d G_0^3 N^{-3+3 \beta } \left(\mathcal{K}(N)\right)^{-3+6 b} \beta ^3
\, ,
\end{align}
with $\mathcal{K}(N)=G_1-G_0 N^{\beta }$ and the corresponding Lagrange multiplier of Eq. (\ref{lagrange}), is equal to,
\begin{align}\label{sder}
& \lambda (N)=-3 \left(\mathcal{K}(N)\right)^{2 b}-216 d \left(\mathcal{K}(N)\right)^{4 b}+432 d \left(\mathcal{K}(N)\right)^{6 b}\\ \notag & +108 b d G_0 N^{-1+\beta } \left(\mathcal{K}(N)\right)^{-1+4 b} \beta -1296 b d G_0 N^{-1+\beta } \left(\mathcal{K}(N)\right)^{-1+6 b} \beta \\ \notag & +972 b^2 d G_0^2 N^{-2+2 \beta } \left(\mathcal{K}(N)\right)^{-2+6 b} \beta ^2-216 b^3 d G_0^3 N^{-3+3 \beta } \left(\mathcal{K}(N)\right)^{-3+6 b} \beta ^3
\, .
\end{align}
Then, by recalling that $\phi=t$, we can in principle find the relation $t=t(\alpha )$, where $\alpha$ is the scale factor, and express the cosmic time as a function of $N$. However, in the present case this is not possible, since the function $t=t(\alpha)$ is of the following form,
\begin{equation}\label{taexplicit}
t=\frac{1}{G_0}F_2\left(b,\frac{1}{\beta },1+\frac{1}{\beta },\frac{G_0 \ln\left(\frac{\alpha }{a_0}\right)^{\beta }}{G_1}\right) \ln\left(\frac{\alpha }{a_0}\right) ^b \, ,
\end{equation}
which cannot be solved as a function of $\alpha$, where $F_2$ is the Gauss hypergeometric function, but the potential (\ref{explicitpot1}) provides enough information for the behavior of the mimetic matter. There are cases however, that it is possible to find the potential as a function of $\phi$, for example for the cosmological evolution with Hubble rate,
\begin{equation}\label{hub2}
H(N)=\left(-G_0\text{  }e^{\beta N }+G_1\right)^b\, ,
\end{equation}
it is possible to find the potential as a function of $\phi$. The slow-roll parameter $\epsilon$ for this evolution reads,
\begin{align}\label{hubslowroll2}
& \epsilon=-\frac{b e^{\beta N} G_0 \beta  \left(G_1 (6+\beta )-2 e^{\beta N} G_0 (3+b \beta )\right)^2}{4 \mathcal{F}(N)}
\end{align}
with the function $\mathcal{F}(N)$, being equal to,
\begin{equation}\label{hfghfhgfghdf}
\mathcal{F}(N)=\left(e^{\beta N} G_0-G_1\right) \left(-3 G_1+e^{\beta N} G_0 (3+b \beta )\right)^2
\end{equation}
while the parameter $\eta$ can be easily calculated and it is equal to,
\begin{equation}\label{edggs}
\eta =-\frac{\beta  \left(8 b^2 e^{2\beta N} G_0^2 \beta +G_1 \left(2 e^{\beta N} G_0 (-3+\beta )+G_1 (6+\beta )\right)+2 b e^{\beta N} G_0 \left(12 e^{\beta N} G_0-G_1 (12+5 \beta )\right)\right)}{4 \left(e^{\beta N} G_0-G_1\right) \left(-3 G_1+e^{\beta N} G_0 (3+b \beta )\right)}
\, .
\end{equation}
By combining Eqs. (\ref{hubslowroll2}) and (\ref{edggs}), we can easily compute the corresponding spectral index and scalar-to-tensor ratio, with $n_s$ being equal to,
\begin{align}\label{scalarpertandsctotenso}
& n_s=\frac{2 \left(e^N\right)^{3 \beta } G_0^3 (3+b \beta )^2 (1+2 b \beta )+3 G_1^3 \left(-6+6 \beta +\beta ^2\right)}{2 \mathcal{F}(N)}
+\frac{e^{\beta N} G_0 G_1^2 \left(54+12 (-3+4 b) \beta +3 \beta ^2+2 b \beta ^3\right)}{2 \mathcal{F}(N)}\\ \notag &
-\frac{2 e^{2\beta N} G_0^2 G_1 \left(27+(-9+48 b) \beta +\left(3+13 b^2\right) \beta ^2+b (1+b) \beta ^3\right)}{2 \mathcal{F}(N)}
\, .
\end{align}
and the scalar-to-tensor ratio is equal to,
\begin{equation}\label{thodorakis}
r=-\frac{4 b e^{\beta N} G_0 \beta  \left(G_1 (6+\beta )-2 e^{\beta N} G_0 (3+b \beta )\right)^2}{\mathcal{F}(N)}
\end{equation}
By choosing the parameters $G_0$, $G_1$, $\beta$ and $b$ as follows,
\begin{equation}\label{parmchoice12}
G_0=0.5,\,\,\, G_1=10,\,\,\,\beta=0.0222,\,\,\, b=1\, ,
\end{equation}
the indices $n_s$ and $r$ read,
\begin{equation}\label{indnewparadigm12}
n_s\simeq 0.96567, \,\,\, r=0.0640848\, ,
\end{equation}
which are in concordance with the Planck data (\ref{constraintedvalues}). Of course there are other values of the parameters $G_0$, $G_1$, $\beta$ and $b$, for which the observational indices are compatible with observations, like for example if we choose, $G_0=0.5$, $G_1=12$, $\beta=0.024$ and $b=1/2$, but for the choice (\ref{parmchoice12}), it is possible to express the potential $V(N)$, as a function of $\phi$. Indeed, we can easily obtain the potential $V(N)$, which is,
\begin{align}\label{podnegsa}
& V(N)=6 \left(\mathcal{S}(N)\right)^{2 b}+12 d \left(\mathcal{S}(N)\right)^{2 b}+3 \left(\mathcal{S}(N)\right)^{4 b}+27 d \left(\mathcal{S}(N)\right)^{12 b}+2 b e^{\beta N} G_0 \left(\mathcal{S}(N)\right)^{-1+2 b} \beta\\ \notag & +432 d \left(\mathcal{S}(N)\right)^{4 b}+1836 d \left(\mathcal{S}(N)\right)^{6 b}
+1296 d \left(\mathcal{S}(N)\right)^{8 b}+324 d \left(\mathcal{S}(N)\right)^{10 b} \, ,
\end{align}
with $\mathcal{S}(N)=-e^{N\beta } G_0+G_1$, and the corresponding Lagrange multiplier reads,
\begin{align}\label{lajdbfe}
& \lambda (N)=-3 \Big{(}\mathcal{S}(N)\Big{)}^{2 b}-6 d \Big{(}\mathcal{S}(N)\Big{)}^{2 b}-\frac{3}{2} \Big{(}\mathcal{S}(N)\Big{)}^{4 b}-18 d \Big{(}\mathcal{S}(N)\Big{)}^{6 b}-18 b d e^{N\beta } G_0 \Big{(}\mathcal{S}(N)\Big{)}^{-1+6 b} \beta\\ \notag & -\frac{9}{2} d \Big{(}\mathcal{S}(N)\Big{)}^{8 b}-3 b e^{N\beta } G_0 \Big{(}\mathcal{S}(N)\Big{)}^{-1+2 b} \beta -72 b d e^{N\beta } G_0 \Big{(}\mathcal{S}(N)\Big{)}^{-1+4 b} \beta
 \, .
\end{align}
By choosing the parameters as in Eq. (\ref{parmchoice12}), it is possible to express the cosmic time as a function of the scale factor $\alpha$, in the following way: since $H(N(\alpha))$ is equal to,
\begin{equation}\label{hna}
H(N(\alpha))=\left(-G_0\text{  }(\alpha /a_0)^{\beta }+G_1\right)^b\, ,
\end{equation}
owing to the fact that $e^N=\frac{\alpha}{\alpha_0}$, by integrating the above equation and for the values of the parameters given in Eq. (\ref{parmchoice12}) ($b=1$ is the most relevant for the calculation), we obtain the function $t=t(\alpha)$, which is,
\begin{equation}\label{ta}
t+c_1=\frac{\beta  \ln(\alpha )-\ln\left(G_1-G_0 \left(\frac{\alpha }{a_0}\right)^{\beta }\right)}{G_1 \beta }\, ,
\end{equation}
which can be solved with respect to the scale factor $\alpha$,
\begin{equation}\label{scalef}
\alpha (t)= \left (\frac{e^{G_1\beta(t+c_1)}G_1}{1+e^{G_1\beta(t+c_1)}\frac{G_0}{\alpha_0^{\beta}}}\right)^{\frac{1}{\beta}},\,\,\,c_1=\frac{1}{\beta G_1}\ln \left(\frac{\alpha_0}{G_1-\frac{G_0}{\alpha_0^{\beta-1}}}\right)\, .
\end{equation}
Recalling that $\phi=t$, by making the following substitution in Eq. (\ref{podnegsa}),
\begin{equation}\label{finalsubstitution}
e^{\beta N} \rightarrow \frac{e^{G_1\beta(\phi+c_1)}G_1}{1+e^{G_1\beta(\phi+c_1)}\frac{G_0}{\alpha_0^{\beta}}}\, ,
\end{equation}
we easily obtain the scalar potential $V(\phi)$. In the context of the reconstruction method we used, it is possible to reproduce a large number of phenomenologically appealing cosmologies, by just using any viable $F(R)$ gravity one may choose and also any given Hubble rate, which produces a cosmological evolution compatible with observational data. We used a simple $F(R)$ function, but in principle one can choose an exponential $F(R)$ gravity  or other viable models \cite{reviews1}. In this way, it is possible that the gravitational theory at hand, may have all the appealing attributes of $F(R)$ theories, like explaining early and late-time acceleration, and at the same time providing an alternative candidate for dark matter.

\section{Inverse Reconstruction Method for Determining the $F(R)$ Gravity}

The reconstruction method we presented in this letter can work in the opposite way too, for example by providing the Hubble rate, the potential $V(\phi )$ and the Lagrange multiplier $\lambda (\phi )$, it is possible to find the $F(R)$ gravity which generates such a cosmological evolution. Let us exemplify this by using a symmetric bounce, which is a very well known bounce cosmology \cite{symmetricbounce}, with the scale factor and the corresponding Hubble rate being equal to,
\begin{equation}\label{symmetricb}
a(t)=e^{\gamma t^2},\, \, \, H(t)=2\gamma t\, .
\end{equation}
Suppose that the potential $V(\phi=t)$ and the Lagrange multiplier $\lambda (t)$ are given as follows,
\begin{equation}\label{potandlag}
V(t)=-24 t^2 \gamma ^2,\, \, \, \lambda (t)=-2\gamma\, .
\end{equation}
Having at hand the cosmological evolution (\ref{symmetricb}), the potential and the Lagrange multiplier of Eq. (\ref{potandlag}), by making use of Eqs. (\ref{scalarpot}) and (\ref{lagrange}), we may easily obtain the $F(R)$ gravity that generates such a cosmological scenario. Upon combining Eqs. (\ref{scalarpot}) and (\ref{lagrange}), we obtain the following differential equation,
\begin{equation}\label{diffeqnnewaddition}
2\frac{\mathrm{d}^2F'(R)}{\mathrm{d}t^2}+4H\frac{\mathrm{d}F'(R)}{\mathrm{d}t}- 6 H \frac{\mathrm{d}F'(R)}{\mathrm{d}t}-2 (\dot{H}+3 H^2)+6(\dot{H}+H^2)\frac{\mathrm{d}F'(R)}{\mathrm{d}t}-2 \lambda (t)-V(t)=0\, \end{equation}
Using Eqs. (\ref{symmetricb}) and (\ref{potandlag}), and applying these to Eq. (\ref{diffeqnnewaddition}), we obtain the following differential equation,
\begin{equation}\label{diifequation}
2\ddot{y}(t)+4 H(t)\dot{y}(t)-6 H(t) \dot{y}(t)+6 \left(\dot{H}(t)+H(t)^2\right)y(t)-2 \left(\dot{H}(t)+3 H(t)^2\right)-2 \lambda (t)-V(t)=0\, ,
\end{equation}
where $y(t)=F'(R(t))$, and  solving this differential equation with respect to $y(t)$, one gets the following solution,
\begin{equation}\label{finalfrsolution}
y(t)=c_1 e^{-\frac{1}{2} \left(5+\sqrt{37}\right) t^2 \gamma } H_{n_1}(37^{1/4} t \sqrt{\gamma })+c_2e^{-\frac{1}{2} \left(5+\sqrt{37}\right) t^2 \gamma }{}_1F_1(-\frac{n_1}{2},\frac{1}{2},\sqrt{37} t^2 \gamma)\, ,
\end{equation}
where $n_1=\frac{1}{74} \left(-37-11 \sqrt{37}\right)$, $c_1$ and $c_2$ arbitrary constants, and the functions $H_{n}(x)$ and ${}_1F_1(a,b,x)$ are the Hermite polynomial and the Kummer confluent hypergeometric function respectively. By also recalling that $R=6 \dot{H}+12 H^2$, substituting the Hubble rate (\ref{symmetricb}) and solving with respect to $t$, we obtain, $t=\frac{\sqrt{R-12 \gamma }}{4 \sqrt{3} \gamma }$, so by substituting this in (\ref{finalfrsolution}), we can easily obtain the function $y(R)$. Then we can obtain the analytic form of $F(R)$ in the large and small $R$ limits. For example, for large $R$, the $F'(R)$ gravity is approximately equal to,
\begin{equation}\label{approxfr}
F'(R)\simeq C_1+C_2 R\, ,
\end{equation}
where $C_1$ and $C_2$ are constants, the analytic form of which is irrelevant to our analysis and can be found in the Appendix A.

\section{Existence of de Sitter Solutions and Dynamical Evolution}

The early time acceleration era is described by (quasi)-de Sitter solutions, so firstly it is important to investigate if de Sitter solutions exist in the case of mimetic $F(R)$ theory with potential $V(t=\phi)$. If a specific $F(R)$ has de Sitter solutions, we need to examine whether these de Sitter solutions are stable against linear perturbations. If the de Sitter solutions are stable, then this solution is the final attractor of the theory and therefore the eternal inflationary era lasts forever\cite{linde1,sergeitracepaper}, and on the contrary, if the perturbation is unstable, the graceful exit from inflation is achieved. We start off with the $F(R)$ gravity of Eq. (\ref{staro}). For this model, de Sitter solutions exist only for $d<0$, as we now demonstrate. The potential $V(t)$ and the Lagrange multiplier $\lambda (t)$, for the Hubble rate being $H(t)=H_{dS}$, take the following form,
\begin{equation}\label{avtadesitte}
V(t)=6 H_{dS}^2-1728 d H_{dS}^6,\,\,\, \lambda(t)=-3 H_{dS}^2-432 d H_{dS}^6\, ,
\end{equation}
and by substituting these to Eq. (\ref{diffeqnnewaddition}), we get the equation $6 H_{dS}^2 \left(-1+432 d H_{dS}^4\right)=0$, and by solving it, with respect to $H_{dS}$, we obtain the only de Sitter solution that exists, which is of the form,
\begin{equation}\label{equationsol}
H_{dS}=\frac{1}{2\ 3^{3/4} d^{1/4}}\, .
\end{equation}
What now remains is to study the linear stability of this solution. An unstable de Sitter solution indicates that the  inflation ends. We consider perturbative solutions of the de Sitter solution $H(t)=H_{dS}$, which have the following form,
\begin{equation}\label{perturbation}
H(t)=H_{dS}+\Delta H(t)\, ,
\end{equation}
with the perturbation being very small, that is $\mid\Delta H(t)\mid\ll 1$. Substituting this to Eq. (\ref{diffeqnnewaddition}), and by keeping terms linear to $\Delta H(t)$ and it's higher derivatives $\Delta \dot{H}(t)$ and $\Delta \ddot{H}(t)$, we obtain the following differential equation for the perturbation $\Delta H(t)$,
\begin{align}\label{ddotdiffeqna}
 -2592 d H_{dS}^4 \Delta \ddot{H}(t)-12 H_{dS} \Delta \dot{H}(t)+15552 d H_{dS}^5 \Delta \dot{H}(t)-6 \Delta \dot{H}(t)+5184 d H_{dS}^4 \Delta \dot{H}(t)&\\ \notag
-10368 d H_{dS}^5 \Delta \dot{H}(t)-6 H_{dS}^2+2592 d H_{dS}^6=0&
\end{align}
where the ``dot'' here denotes differentiation with respect to the cosmic time $t$. Note that we used again Eqs. (\ref{scalarpot}) and (\ref{lagrange}) to which we substituted Eq. (\ref{perturbation}). The solution to this equation is of the form,
\begin{equation}\label{solutionbigfr3}
\Delta H(t)=\mathcal{C}+\mathcal{C}_1e^{\zeta_1t}+\mathcal{C}_2e^{\zeta_2 t}\, ,
\end{equation}
where $\mathcal{C}_1,\mathcal{C}_2$ are arbitrary integration constants and the exact analytic form of the parameters $\mathcal{C}$, $\zeta_1$ and $\zeta_2$ can be found in the Appendix, since these are too complicated to be presented here. By substituting the exact de Sitter value $H_{dS}$ of Eq. (\ref{equationsol}), in the parameter $\zeta_2$, we can see that the parameter $\zeta_2$ is positive, so the term $\sim e^{\zeta_2 t}$ dominates the evolution, since the other term contains $\zeta_1$, which is negative. Therefore, the de Sitter solution we found in the $F(R)=R-dR^3$ model is unstable, and therefore inflation can end in this model, always in the presence of mimetic potential and Lagrange multiplier.

Before closing our study, it is worth presenting the results of another model, for which a de Sitter solution exists and the perturbation evolution is much more unstable in comparison to the model we just studied. The model is described by the following $F(R)$ function,
\begin{equation}\label{frfunctoim}
F(R)=R-d R^3+f R^2\, ,
\end{equation}
so it results from the model of Eq. (\ref{staro}), by simply adding an $R^2$ term. Interestingly enough, the model (\ref{frfunctoim}) is known to yield quite interesting solutions for compact objects as was shown in \cite{sergeicappostars}. The intriguing feature is that when the $F(R)$ gravity is taken exactly as in Eq. (\ref{frfunctoim}) (that is, the coefficient of $R^3$ negative and the coefficient of $R^2$ positive), there appears an increase in the maximal neutron star mass. The full study of the model (\ref{frfunctoim}) will be presented elsewhere, here we describe only the results of our analysis. The de Sitter solution for the model (\ref{frfunctoim}) is the following,
\begin{equation}\label{desitterr2r3}
H_{dS}=\frac{1}{6} \sqrt{\frac{f}{d}+\frac{\sqrt{3 d+f^2}}{d}}\, ,
\end{equation}
and the evolution of linear perturbations behaves as,
\begin{equation}\label{sevolutggrees}
\Delta H(t)=\mathcal{A}+c_Ae^{\delta_1 t}+c_Be^{\delta_2 t}\, ,
\end{equation}
where $\mathcal{A}$, $\delta_1$ and $\delta_2$ appear in the Appendix, and the parameters $c_A$ and $c_B$ are integration constants. The parameter $\delta_1$ is negative, and the parameter $\delta_2$ is positive. In addition, $\delta_2$ grows larger as $f$ and $d$ appearing in Eq. (\ref{frfunctoim}) increase. The resulting picture is that the linear perturbations of the de Sitter solution of the model (\ref{frfunctoim}) are much more unstable in comparison to the model (\ref{staro}), which means that the $R^2$ term makes the model more unstable towards linear perturbations of the de Sitter solutions as it was demonstrated long ago by Starobinsky \cite{star}. The detailed study of this model will be given elsewhere. Finally, we need to mention that the model $F(R)=R+d R^3+f R^2$ has no de Sitter solutions, so only the model with negative $d$ and it's $R^2$ deformations have de Sitter solutions.

\section{Conclusion}

In this letter we studied the behavior of mimetic $F(R)$ gravity with scalar potential. As we demonstrated it is possible to generate inflationary cosmology consistent with observations by using specific forms of the mimetic potential and Lagrange multiplier, for a given form of the Hubble rate. After exemplifying our findings by using some characteristic examples, we investigated the existence of de Sitter solutions for the  mimetic potential $F(R)$ theory and also we examined the stability of these de Sitter solutions. As we evinced, the final result strongly depends on the form of the $F(R)$ gravity. It is worth to study the existence of de Sitter solutions for specific classes of potentials. This procedure would narrow down the possible $F(R)$ gravities that would yield de Sitter solutions for the given potential.

In addition, the Einstein frame counterpart theory of the mimetic $F(R)$ theory can easily be found, and the final form of the resulting two scalar theory is interesting to study. Particularly, since in the Einstein frame there is a direct coupling between the Einstein frame scalar and the mimetic scalar, it would be interesting to investigate the resulting theory, which can have similarities with some recently studied two scalar field models \cite{linde}, if the Einstein frame scalar counterpart of the $F(R)$ gravity is chosen to be the $R^2$ inflation model \cite{star,starobinskyoldpot}. Also the issue of gravitational particle production should be re-addressed in such a theoretical framework.

Moreover, it would be very interesting to reproduce known consistent with observations Jordan frame cosmologies, such as the $R^2$ inflation cosmology, by using an arbitrary viable $F(R)$ gravity and a suitably chosen mimetic potential. Finally, a study of the dynamical system that corresponds to the $F(R)$ gravity with mimetic potential would be of importance. Specifically, it is worth investigating which are the fixed points of the theory, also perform a stability analysis of these points, and also providing the physical meaning of some of these fixed points could be of importance, since the mimetic potential could have interesting implications on the physical meaning of the fixed points. Finally, we need to note that in this paper we assumed a kind of slow-roll approximation in order to extract the observational indices in the Jordan frame. It is possible however to abandon the slow-roll approximation and to study different inflation regimes (compare with \cite{barrowslowroll}).

Finally, we need to note that the mimetic $F(R)$ formalism can be applied in such a way so contact with the present Universe is achieved. Particularly, this can be done by using some recent cosmographic developments \cite{cosmo1,cosmo2}. There are three possible ways to accommodate cosmographic data in the mimetic $F(R)$ formalism.  Firstly, one could use the resulting Hubble rate of cosmographic data, which is of the form \cite{cosmo1},
\begin{equation}\label{capowner}
H(z)\sim \sqrt{\Omega_m(1+z)^3+\ln (\alpha+\beta z)}
\end{equation}
where $z$ is the redshift, and $\alpha$ and $\beta$ parameters of the theory (see \cite{cosmo2} for details), and for a general $F(R)$ gravity it is possible to find the Lagrange multiplier $\lambda (t)$ and the potential $V(t)$ that can yield the evolution (\ref{capowner}). The second way that contact with cosmography can be achieved, is to use the $F(R)$ gravity which corresponds to the latest cosmographic data \cite{cosmo2} (see Eq. 31 of \cite{cosmo2}), and find the potential and Lagrange multiplier that generates such a cosmological evolution. Finally, the third way to make contact with the latest cosmographic data is to use both the Hubble rate and the $F(R)$ gravity that correspond to cosmographic data, and again investigate how the potential and the Lagrange multiplier behave for these choices. Lastly, let us note that it is quite interesting to generate the cosmological evolution of Eq. (\ref{capowner}) by using the mimetic $F(R)$ gravity formalism, with the $F(R)$ gravity being slightly different from the General Relativity case, for example $F(R)\sim R^{1+\epsilon}$, with $\epsilon \ll 1$. In this way we could find a power expansion of the Lagrange multiplier and of the potential, in terms of the parameter $\epsilon$ and the resulting expressions would be quite simple. We hope to address these issues in a future work.

\section*{Acknowledgments}

This work is supported in part by MINECO (Spain), project FIS2013-44881 (S.D.O) and Min.of Education and Science of Russia (S.D.O and V.K.O.).

\section*{Appendix: Detailed Expressions of Parameters Appearing in Text}

Here we quote the exact form of the parameters $C_1$, $C_2$, $\mathcal{C}$, $\zeta_1$, $\zeta_2$, $\mathcal{A}$, $\delta_1$ and $\delta_2$ appearing in the main text of the paper. We start off with $C_1$ and $C_2$ which are the coefficients of the small curvature expansion of the $F'(R)$ gravity, appearing in Eq. (\ref{approxfr}), the analytic form of which is,
\begin{align}\label{alc1c2}
& C_1=\left(c_1 e^{\frac{1}{8} \left(5+\sqrt{37}\right)} H_{n_a}\left(,\frac{37^{1/4} \sqrt{-\gamma }}{2 \sqrt{\gamma }}\right)+c_2 e^{-\frac{1}{4} \left(5+\sqrt{37}\right) \sqrt{-\gamma }} {}_1F_1\left(-\frac{n_a}{2},\frac{1}{2},-\frac{\sqrt{37}}{4}\right)\right),\\ \notag &
C_2=\Big{(}c_1 \left(\frac{\left(-5-\sqrt{37}\right) e^{\frac{1}{8} \left(5+\sqrt{37}\right)} H_{n_a}\left(\frac{37^{1/4} \sqrt{-\gamma }}{2 \sqrt{\gamma }}\right)}{96 \gamma }+\frac{\left(37+11 \sqrt{37}\right) e^{\frac{1}{8} \left(5+\sqrt{37}\right)} \sqrt{-\gamma } H_{-1+n_a}\left(\frac{37^{1/4} \sqrt{-\gamma }}{2 \sqrt{\gamma }}\right)}{48\ 37^{3/4} \gamma ^{3/2}}\right)\\ \notag & +c_2 \left(\frac{\left(5+\sqrt{37}\right) e^{-\frac{1}{4} \left(5+\sqrt{37}\right) \sqrt{-\gamma }} \sqrt{-\gamma } {}_1F_1\left(-\frac{n_a}{2},-\frac{\sqrt{37}}{4}\right)}{96 \gamma }+\frac{\left(37+11 \sqrt{37}\right) e^{-\frac{1}{4} \left(5+\sqrt{37}\right) \sqrt{-\gamma }} {}_1F_1\left(1-\frac{n_a}{2},\frac{3}{2},-\frac{\sqrt{37}}{4}\right)}{96 \sqrt{37} \gamma }\right)\Big{)}\, ,
\end{align}
where $n_a$ is equal to,
\begin{equation}\label{na}
n_a=\frac{1}{74} \left(-37-11 \sqrt{37}\right)\, .
\end{equation}
The parameters $\mathcal{C}$, $\zeta_1$ and $\zeta_2$ that appear in Eq. (\ref{solutionbigfr3}), are equal to,
\begin{align}\label{newparameters}
& \mathcal{C}=\frac{H_{dS} \left(-1+432 d H_{dS}^4\right)}{2 \left(1-1296 d H_{dS}^4\right)}\, , \\ \notag &
\zeta_1=\frac{\left(-1+864 d H_{dS}^4-1728 d H_{dS}^5-\sqrt{1-1728 d H_{dS}^4+746496 d^2 H_{dS}^8+1492992 d^2 H_{dS}^9+2985984 d^2 H_{dS}^{10}}\right) }{864 d H_{dS}^4}\, , \\ \notag &
\zeta_2=\frac{\left(-1+864 d H_{dS}^4-1728 d H_{dS}^5+\sqrt{1-1728 d H_{dS}^4+746496 d^2 H_{dS}^8+1492992 d^2 H_{dS}^9+2985984 d^2 H_{dS}^{10}}\right) }{864 d H_{dS}^4}\, .
\end{align}
Finally, the parameters $\mathcal{A}$, $\delta_1$ and $\delta_2$ appearing in Eq. (\ref{sevolutggrees}), are equal to,
\begin{align}\label{newparametersdeltaparameters}
& \mathcal{A}=\frac{H_{dS} \left(-1-24 d H_{dS}^2+432 d H_{dS}^4\right)}{2 \left(1+48 d H_{dS}^2-1296 d H_{dS}^4\right)}\, , \\ \notag &
\delta_1=\frac{1}{2 \left(-12 d H_{dS}^2+432 d H_{dS}^4\right)}\Big{(}-1-36 d H_{dS}^2+48 d H_{dS}^3+864 d H_{dS}^4-1728 d H_{dS}^5\\ \notag & -\sqrt{-4 \left(-12 d H_{dS}^2+432 d H_{dS}^4\right) \left(2 H_{dS}+96 d H_{dS}^3-2592 d H_{dS}^5\right)+\left(1+36 d H_{dS}^2-48 d H_{dS}^3-864 d H_{dS}^4+1728 d H_{dS}^5\right)^2}\Big{)}\, , \\ \notag &
\delta_2=\frac{1}{2 \left(-12 d H_{dS}^2+432 d H_{dS}^4\right)}\Big{(}-1-36 d H_{dS}^2+48 d H_{dS}^3+864 d H_{dS}^4-1728 d H_{dS}^5\\ \notag & +\sqrt{-4 \left(-12 d H_{dS}^2+432 d H_{dS}^4\right) \left(2 H_{dS}+96 d H_{dS}^3-2592 d H_{dS}^5\right)+\left(1+36 d H_{dS}^2-48 d H_{dS}^3-864 d H_{dS}^4+1728 d H_{dS}^5\right)^2}\Big{)}\, .
\end{align}

\end{document}